\documentclass[prb,twocolumn,showpacs,amsmath,amssymb,superscriptaddress]{revtex4}

\usepackage{graphicx}

\newcommand{\ix}[1]{\ensuremath{\text{\textsl{#1}}}}
\newcommand{\abs}[1]{\ensuremath{\left| #1 \right|}}
\newcommand{\agm}[2]{\ensuremath{\text{agm} \left(#1, #2 \right)}}

\begin{document}

\title{Temperature induced phase averaging in one-dimensional mesoscopic systems}
\author{Severin~G.~Jakobs}
\affiliation{Institut f\"ur Theoretische Physik~A, RWTH Aachen, D-52056
  Aachen, Germany}
\author{Volker~Meden}
\affiliation{Institut f\"ur Theoretische Physik, Universit\"at G\"ottingen,
  D-37077 G\"ottingen, Germany}
\author{Herbert~Schoeller}
\affiliation{Institut f\"ur Theoretische Physik~A, RWTH Aachen, D-52056
  Aachen, Germany}
\author{Tilman~Enss}
\affiliation{Dipartimento di Fisica, Universit\`a di Roma ``La Sapienza'',
  Piazzale Aldo Moro 2, I-00185 Roma, Italy}
\date{\today}

\begin{abstract}
  We analyse phase averaging in one-dimensional interacting 
  mesoscopic systems with several barriers and show that 
  for incommensurate positions an independent average over several 
  phases can be induced by finite temperature.
  For three strong barriers with conductances $G_i$ and mutual distances 
  larger than the thermal length, we obtain $G\sim\sqrt{G_1G_2G_3}$ for the total
  conductance $G$. For an interacting wire, this
  implies power laws in $G(T)$ with novel exponents, which we propose
  as an experimental fingerprint 
  to distinguish temperature induced phase averaging from dephasing.
\end{abstract}


\maketitle

\section{Introduction}
Mesoscopic systems are characterized by spatial dimensions smaller
than the phase breaking length $L_\varphi$ so that the phase of an
electron is not destroyed by inelastic processes. Along their optical
paths, electrons pick up phases from propagation and scattering. Even
for negligible dephasing, certain circumstances can lead to an 
\emph{averaging} of these phases, a phenomenon which has been
analyzed in connection with localization, see 
e.g.\ Refs.~\onlinecite{gang4, imry_shiren}. 
It remains however a fundamental task to compare in detail 
the effects of phase averaging and dephasing.
For a noninteracting model system it was recently shown 
that a single one-channel dephasing probe gives the same
full counting statistics as a single phase-averaging 
probe.\cite{pilgram_etal} However, it was noted
that this result does no longer hold for
several probes, i.e.\ phase averaging over many independent
phases seems to be fundamentally different from dephasing.
In the present paper we address this issue in detail. We discuss a generic
one-dimensional interacting system coupled to two reservoirs 
with few barriers at arbitrary but fixed positions.
We show that for incommensurate barrier positions an independent
average over several phases can be induced by finite temperature.
For more than two barriers we find that phase averaging drastically 
differs from dephasing and show that the commonly made assumption that 
barriers with mutual distances larger than the thermal length
$L_T=v_\ix{F}/T$ can be treated independently
\cite{egger_gogolin,LL_exp,postma} (note that in experiment
contacts to leads constitute additional barriers) 
does not hold in the absence of dephasing. This means that even in the
high-temperature regime interference effects are still important, as also
shown recently for quantum dots in the sequential tunneling
regime.\cite{koenig_gefen,koenig_braun}
For three strong barriers with individual conductances $G_i$, we obtain
$G \sim \sqrt{G_1G_2G_3}$ for the total conductance, in drastic contrast to 
the addition of resistances which follows from dephasing.
For an interacting quantum wire (QW) this implies power-law
scaling of $G(T)$ with novel exponents which can be used as an 
experimental fingerprint for phase averaging.

In section \ref{sec:noninteracting} we introduce a noninteracting
model for a quantum wire with barriers which allows a simple
discussion of temperature induced phase averaging. In section
\ref{sec:interacting} we study the influence of temperature induced
phase averaging on the scaling beghaviour of interacting quantum wires
and propose an experimental setup to distinguish phase averaging from
dephasing. While sections  \ref{sec:noninteracting} and
\ref{sec:interacting} focus primarily on a wire with three barriers, section
\ref{sec:4barrier} treats phase averaging for a wire with four barriers.

\section{Noninteracting wire}
\label{sec:noninteracting}
\begin{figure}[tbh]
  \includegraphics[width=0.9\linewidth]{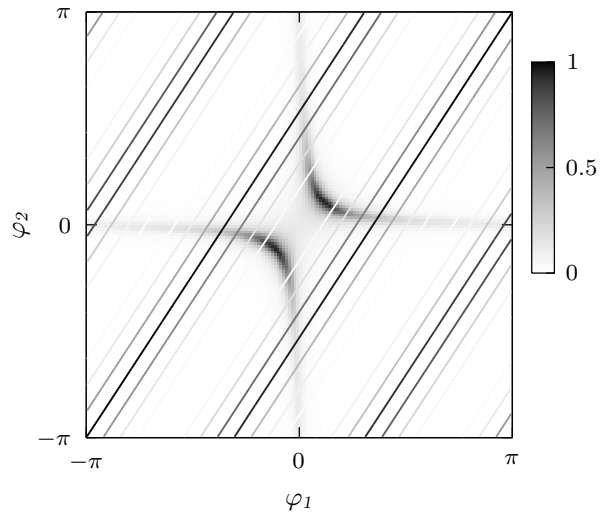}
  \caption{
    Density plot of the transmission probability $\abs{t(\varphi_1,
      \varphi_2)}^2$ as function of the phases $\varphi_{1,2}$ for a 
    noninteracting QW with three barriers $\tau_{1,2,3} = 0.35$.
    Lines: Path which the integral in Eq.~\eqref{eq:landauer_buettiker}
    traces in the $(\varphi_1,\varphi_2)$-plane for $N_1 = 4940$, $N_2
    = 7560$. The grayscale of the line indicates the relative amplitude of
    $-\partial f/ \partial \epsilon$ for $T = 1.5 \cdot 10^{-3}$, the same
    grayscale applies as for the density plot. For sufficiently high
    temperature the line covers a representative part of the plane.
  }
  \label{fig:phase_averaging}
\end{figure}

For a discussion of the basic physical idea
consider first the noninteracting case. We model
the QW coupled to leads by an infinite tight-binding chain
for spinless electrons at half-filling ($\mu=0$)
\begin{equation}
  H=-\sum_{n \in \mathbb{Z}, n\ne n_i} c^\dagger_{n+1} c_{n}
  -\sum_{i=1}^p \tau_i \,c^\dagger_{n_i+1} c_{n_i} + \text{h.c.}
\end{equation}
with hopping matrix elements equal to 1 (defining the energy unit) except
for $n=n_i$, $i=1,\dots,p$, where the barriers are situated. Without
barriers the dispersion relation is $\epsilon=-2\cos{(ka)}$
with band width $D=4$, where $k$ is the momentum and
$a$ the lattice spacing. The distance between subsequent barriers is
$L_i=aN_i$ with $N_i=n_{i+1}-n_i$. We characterize the barriers by
their transmission $t_i(\epsilon)$ and reflection $r_i^\pm(\epsilon) =
|r_i(\epsilon)|e^{i\delta_i^\pm(\epsilon)}$ amplitudes for right- and
left-running scattering waves. The linear conductance in units of
$e^2/h$ follows from \cite{scattering}
\begin{equation}
  \label{eq:landauer_buettiker}
  G(T) = \int \! d \epsilon \left(-
    \frac{\partial f}{\partial \epsilon} \right)
  \abs{t(\epsilon)}^2 \,\,.
\end{equation}
where $f = 1/(e^{\epsilon/T} + 1)$, and $t(\epsilon)$
is the transmission amplitude. Thus, the effect of temperature is an average
of the transmission probability $\abs{t}^2$ over an energy range
$\Delta\epsilon\sim T$. 
This is fundamentally different from dephasing 
which destroys the phase information 
for each individual path of the electron. If one takes account of
dephasing by summing up the classical probabilities of all paths,
one obtains $ |r|^2/|t|^2 = \sum_i |r_i|^2/|t_i|^2$, with
$|r|^2=1-|t|^2$.\cite{datta} Since $|r|^2/|t|^2$ is proportional to
the resistance of a barrier, 
this gives the classical law of adding up resistances in series.
In contrast, we now show that temperature induced
energy averaging leads to completely different results. Since $|t_i(\epsilon)|$ and
$|r_i(\epsilon)|$ vary  slowly with energy on the scale $T\ll D$, temperature
averages only over the rapidly varying phases an electron acquires by
bouncing back and forth between two subsequent barriers
$\varphi_i(\epsilon) = 2k(\epsilon)L_i + \delta_i^-(\epsilon) +
\delta_{i+1}^+(\epsilon)$. As $\delta_i^{\pm}$ depends weakly on
energy, the change of $\varphi_i$ over the energy range
$\Delta\epsilon\sim T$ near the Fermi energy is given by
$\Delta\varphi_i \sim \Delta k L_i \sim
\frac{\Delta\epsilon}{\partial\epsilon/\partial k} L_i \sim L_i/L_T$. 
Thus, for $L_T \sim L_i$, $\Delta\varphi_i$ is roughly given by $2\pi$. For all
$L_T \ll L_i$ and incommensurate length $L_i$ (i.e., $m_i L_j = m_j
L_i$ is only valid for large coprime integers $m_{i,j}$), this implies
that the average over energy will cover a representative part of
the multidimensional phase space defined by
$(\varphi_1,\dots,\varphi_{p-1})$ (see
Fig.~\ref{fig:phase_averaging}). In this case, temperature induced 
energy averaging is equivalent to an
\emph{independent} average over all phases $\varphi_i$, and we can replace $|t|^2$ in
Eq.~\eqref{eq:landauer_buettiker} by the phase-averaged quantity 
\begin{equation}
  \label{eq:phase_average}
\langle |t|^2 \rangle = \prod_{i=1}^{p-1} \frac{1}{2\pi}
  \int_0^{2\pi} \! d \varphi_i \, |t|^2
\end{equation}
provided that $v_\ix{F}/L_i \ll T \ll D$. 
Note that this result does
not hold for commensurate lengths $L_i$ (where the
path in phase space closes very quickly; see below) and becomes less
relevant for a large number $p$ of barriers (since $L_i/L_T$
has to be chosen too large in order to
cover a representative part of the high dimensional phase space).

\begin{figure}[tbh]
  \includegraphics[width=0.9\linewidth]{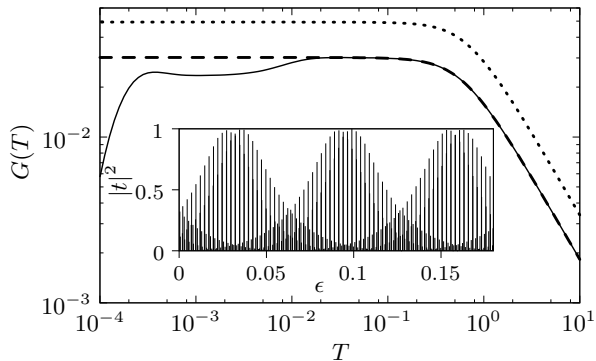}
  \caption{
    Linear conductance as function of temperature
    for a noninteracting QW with three
    barriers $\tau_{1,2,3} = 0.2$ and incommensurate
    dot length, $N_\ix{L}= 4980$, $N_\ix{R} = 7520$. 
    The solid line corresponds to Eq.~\eqref{eq:landauer_buettiker}. For higher $T$,
    the dashed line which results from using 
    the phase averaged transmission given by
    Eq.~\eqref{eq:3barrier} in
    Eq.~\eqref{eq:landauer_buettiker} matches perfectly. Summing up
    the individual resistances (dotted line)
    yields an incorrect result. \emph{Inset: } $\abs{t}^2$ as a
    function of energy.
    The changing mutual energetic distance of resonances of the left
    and right dot leads to energetically large superstructures of the
    total transmission probability.
  }
  \label{fig:nonint_incommens}
\end{figure}

The phase average can be calculated easily for a small number of barriers.
For two barriers we use $t=t_1t_2/(1 - e^{i\varphi_1}|r_1r_2|)$ 
and for three barriers (forming two dots which we refer to as left and
right)
\begin{eqnarray}
  t &=& t_1 t_2 t_3 \big[ \left(1 - e^{i \varphi_\ix{L}} \abs{r_1
      r_2}\right) \left(1 - e^{i \varphi_\ix{R}} \abs{r_2
      r_3}\right) \nonumber\\
  && \hspace{18ex} + e^{i(\varphi_\ix{L} + \varphi_\ix{R})} \abs{r_1
    r_3}\abs{t_2}^2 \big]^{-1}\,,
  \label{eq:t}
\end{eqnarray}
with $\varphi_\ix{L,R} \equiv\varphi_{1,2}$.
By accident, phase averaging yields the addition of resistances 
for two barriers. In contrast, for three barriers we obtain 
\begin{equation}
  \label{eq:3barrier}
  \langle |t|^2 \rangle =\frac{T_1 T_2 T_3}{\sqrt{(\sum_{i<j}T_iT_j)^2
      + 4T_1T_2T_3(1-\sum_i T_i)}}\,\,,
\end{equation} 
with $T_i=|t_i|^2$. For weak barriers, $T_i \approx 1$, this gives
$\langle |t|^2 \rangle \approx T_1T_2T_3$ leading to exponentially small
transmission when scaling it up to many barriers, in analogy to Ref.~\onlinecite{gang4}.
For strong barriers, $T_i\ll 1$, we obtain the surprising
result
\begin{equation}
  \label{eq:3barrier_strong}
  \langle |t|^2 \rangle \approx \frac{1}{2}\sqrt{T_1T_2T_3},
\end{equation}
leading to $G\sim\sqrt{G_1G_2G_3}$ when inserted in Eq.~\eqref{eq:landauer_buettiker}.
Fig.~\ref{fig:nonint_incommens} shows that the phase-averaged
$G$ through three barriers at incommensurate positions agrees 
precisely with the exact one for sufficiently large
$T$, whereas summing up the individual resistances
is incorrect for all $T$. In the inset of Fig.~\ref{fig:nonint_incommens} 
we present the energy dependence of
the total transmission probability. 
It forms energetically large superstructures, as the height and area of the peaks are very
sensitive to the mutual distance of the resonance positions of the
left and right dot. Those are roughly given by even (odd) multiples of
$\pi / (N_\ix{L,R} + 1)$ for odd (even) $N_\ix{L,R}$. Thus, if
$m_\ix{R} (N_\ix{L} + 1) = m_\ix{L} (N_\ix{R} + 1)$ is valid only for
large coprime integers $m_\ix{L,R}$ (incommensurate case
\cite{comment_1}), the mutual energetic distance of left- and right-dot resonance peaks of 
$\abs{t}^2$ shifts slowly along the energy axis, leading to a modulation
of the peak height and area. This provides another picture for 
temperature induced energy averaging:
If $T$ is large enough to average over a sufficient
part of a superstructure, it averages over many mutual distances between
left- and right-dot resonances, which is equivalent to phase averaging.

\begin{figure}[tbh]
  \includegraphics[width=0.9\linewidth]{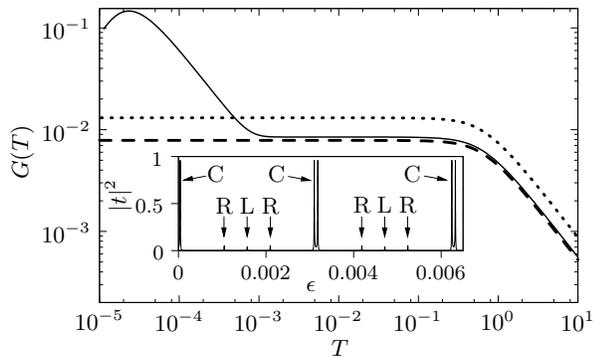}
  \caption{
    Linear conductance as function of temperature
    for a noninteracting
    QW with three barriers $\tau_{1,2,3} = 0.1$ and
    commensurate dot length, $N_\ix{L} + 1 = 4000$, 
    $N_\ix{R} + 1 = 6000$,
    hence $m_\ix{L} = 2, m_\ix{R} = 3$. The solid line results from
    Eq.~\eqref{eq:landauer_buettiker}. For 
    temperatures greater than the distance of coinciding peaks
    (compare inset), the dashed line
    corresponding to Eq.~\eqref{eq:comm_cond} is a valid
    approximation. Summing up the individual resistances (dotted line)
    yields an incorrect result. \emph{Inset: } $\abs{t}^2$ as function
    of energy. Labels indicate well-separated left-dot (L) and
    right-dot (R) resonances as well as coinciding (C) ones.
    The distance of the coinciding resonances is
    $\Delta \epsilon_\text{C} \approx \frac{2 \pi m_\ix{L}}{N_\ix{L}+1} =
    \frac{2 \pi m_\ix{R}}{N_\ix{R}+1} \approx \frac{\pi}{1000}$.
  }
  \label{fig:nonint_commens}
\end{figure}
In contrast, for dots of commensurate length, i.e., when
$m_\ix{R} (N_\ix{L} + 1) = m_\ix{L} (N_\ix{R} + 1)$ is fulfilled 
for small coprime integers $m_\ix{L,R}$, every $m_\ix{L}$th
left-dot and $m_\ix{R}$th right-dot resonance coincide,
whereas the other resonances are
well separated in energy (see inset in Fig.~\ref{fig:nonint_commens}). 
It can be shown that coinciding resonances (avoided crossings) give
rise to two peaks of the total transmission probability, each one with
area $A = 4 \pi \big[(N_\ix{L} + 1)/ \tau_1^2 + (N_\ix{R} +
1)/\tau_3^2\big]^{-1} = {\mathcal O} \left( \tau^2 \right)$.
Note that while $\tau_2$ is proportional to the distance of the two
peaks, it does not influence their area.
The well-separated peaks can be neglected when calculating the total
conductance through the QW because they have a much smaller area of
${\mathcal O} \left(\tau^4 \right)$ than the 
coinciding peaks and they are not in great majority (as $m_\ix{L,R}$ are small). 
For $T \gtrsim 2 \pi \frac{m_\ix{L}}{N_\ix{L} + 1} = 2 \pi
\frac{m_\ix{R}}{N_\ix{R} + 1}$, the energy-integral in
Eq.~\eqref{eq:landauer_buettiker} averages over the 
large coinciding peaks. For the total conductance it follows
(see Fig.~\ref{fig:nonint_commens})
\begin{equation}
  \label{eq:comm_cond}
  \frac{1}{G} = \frac{m_\ix{L}}{G_1} + \frac{m_\ix{R}}{G_3}
\end{equation}
with $G_i = 4 \tau_i^2$. Since the smaller peaks have been
neglected, this formula slightly underestimates
$G$, but becomes more and more accurate for stronger
barriers. Eq.~\eqref{eq:comm_cond} in obviously inconsistent with
summing up the single resistances, and the total conductance is independent of
the strength of the barrier in the middle.

\section{Interacting wire}
\label{sec:interacting}
We now turn to the more realistic case 
of an interacting QW (Luttinger liquid, LL) on sites $n=1,\dots,N$ by adding a 
short ranged interaction 
$H_\ix{\text int}=\sum_{n=1}^{N-1}U_n\rho_n\rho_{n+1}$ with
$\rho_n=c_n^\dagger c_n -\frac{1}{2}$.
Abrupt contacts to the leads follow if the interaction is chosen to be
the same on all sites, $U_n \equiv U$. For smooth contacts $U_n$ rises
slowly from zero at the leads to its full value within the wire over
roughly $100$~sites. The interaction is treated using the truncated functional
renormalization group.\cite{Tilman, meden}
This approach leads to a  real and frequency independent self-energy,
which then serves as an effective single-particle potential so 
that Eq.~\eqref{eq:landauer_buettiker} is still applicable. Inelastic 
processes mediated by the interaction are neglected as they would just
contribute subleading corrections.

A single strong impurity in a LL generates a slowly decaying
oscillatory effective potential of range $L_T$ which 
makes the local density of states in the vicinity of 
a large impurity scale as $\rho(\epsilon = 0) \sim
T^{\alpha_\ix{B}}$.\cite{kane_fisher,glazman_matveev} The boundary
exponent $\alpha_\ix{B}$ can be 
computed from the LL parameter $K$ which in turn is known from Bethe ansatz
\cite{bethe}, $\alpha_\ix{B} = \frac{1}{K} - 1 = \frac{2}{\pi}
\arcsin(\frac{U}{2})$ at half-filling. For $U=1$, as taken in the
following, our approach is known to produce the exponent $\alpha_\ix{B}'
\approx 0.35$ in good agreement with $\alpha_\ix{B} =
1/3$.\cite{Tilman}

If a strong barrier is placed within the interacting part of the
QW, the density of states scales with $T$ on both sides of the
barrier yielding a power-law
$G \sim T^{2 \alpha_\ix{B}}$. If a
strong barrier separates an interacting part of the wire from a
noninteracting part as do tunneling barriers to the leads, $G \sim
T^{\alpha_\ix{B}}$. We are interested in possible power laws if three
barriers of mixed types are present.

\begin{figure}[tb]
  \includegraphics[width=\linewidth]{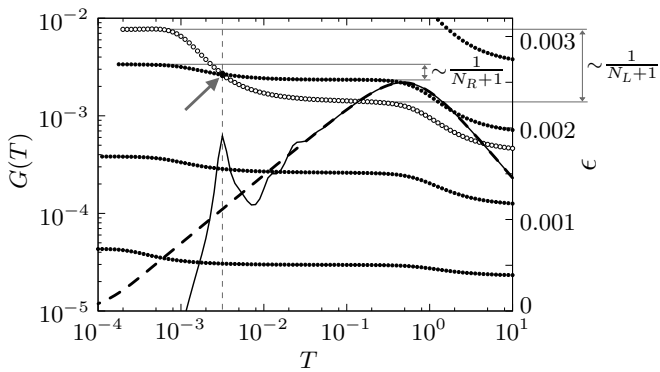}
  \caption{
    $G(T)$ (lines) and position of resonance energies of $\abs{t(\epsilon)}^2$ (circles) 
    for $U=1$ and barriers at the two contacts and a barrier within the wire, 
    $\tau_{1,2,3} = 0.1$, $N_\ix{L} = 1800$, $N_\ix{R} = 8200$.
    The solid line corresponds to Eq.~\eqref{eq:landauer_buettiker}. For
    higher temperatures, the dashed line which results from using
    the phase averaged transmission given by
    Eq.~\eqref{eq:3barrier} in Eq.~\eqref{eq:landauer_buettiker} matches
    perfectly. As left-dot resonances (open circles) are shifted a
    different amount in energy with changing temperature than
    right-dot resonances (filled circles), an avoided crossing can
    occur (arrow)
    leading to a drastically enhanced conductance if temperature is
    low.
  }
  \label{fig:int_incommens}
\end{figure}

\begin{figure}[tb]
  \includegraphics[width=0.9\linewidth]{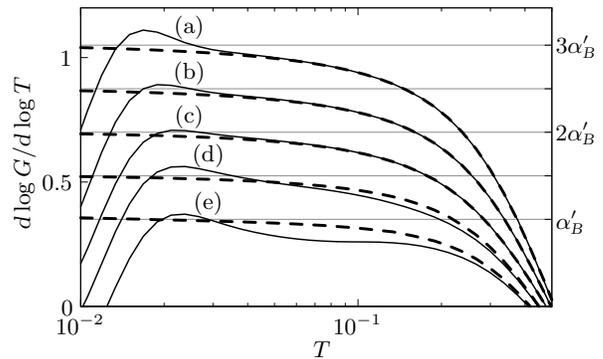}
  \caption{
    Local exponent of $G(T)$ for
    $U=1$ and three barriers, $\tau_{1,2,3} = 0.1$, $N_\ix{L} =
    4930$, $N_\ix{R} = 5070$. (a) All three barriers within the wire.
    (b) One barrier at end-contact to
    lead, two barriers within the wire. (c) Two barriers
    at end-contacts, one barrier within the wire.
    (d) One barrier at bulk-contact, one barrier
    at end-contact, one barrier within the wire.
    (e) Two barriers at bulk-contacts, one barrier within the wire.
    The solid lines correspond to
    Eq.~\eqref{eq:landauer_buettiker}, the dashed lines result from using
    the phase averaged transmission given by
    Eq.~\eqref{eq:3barrier} in Eq.~\eqref{eq:landauer_buettiker}. The
    different levels of the exponents (multiples of $\alpha_\ix{B}'/2$) are
    clearly visible. In the bulk-contacted cases (d) and (e), phase averaging
    is not as good an approximation as in the
    other cases.
  }
  \label{fig:int_incommens_exp}
\end{figure}

Given three barriers in the wire, we find that the effective potential 
formed at one barrier is not affected by the presence of the other barriers,
as long as $L_i \gg L_T$. Consequently, the scaling
law of the transmission probabilities of the individual barriers
$\abs{t_i}^2 \sim \max \{T, \abs{\epsilon}\}^{\alpha_i}$
(with $\alpha_i = \alpha_\ix{B}$ or $2 \alpha_\ix{B}$
depending on the type of the barrier) is not
affected by the presence of the other barriers
and thus $G_i(T)\sim T^{\alpha_i}$. Interpreting a barrier and its surrounding
oscillations as an effective barrier, we can use Eq.~\eqref{eq:t} and 
the picture of two dots in series developed for
the noninteracting case. Thus, 
for incommensurate $L_i$ and sufficiently large $T$, phase averaging
as given by 
Eq.~\eqref{eq:phase_average} can as well be applied to an interacting wire (see
Fig.~\ref{fig:int_incommens}).
This implies power-law scaling of $G \sim \sqrt{G_1 G_2 G_3}$
where the exponent depends on \emph{all} exponents of the individual
barriers
\begin{equation}
  \label{eq:phase_average_scaling}
  G(T) \sim T^{\frac{1}{2}\left(\alpha_1 + \alpha_2 + \alpha_3 \right)}.
\end{equation}
Therefore, combinations of different barriers will produce
exponents equal to different integer multiples of
$\alpha_\ix{B}/2$. For three barriers within the interacting part
of the wire ($\alpha_{1,2,3} = 2 \alpha_\ix{B}$),
Eq.~\eqref{eq:phase_average_scaling} yields $G \sim T^{3
  \alpha_\ix{B}}$. If one of the three barriers is a contact to a
noninteracting lead, it follows $G \sim T^{5 \alpha_\ix{B}/2}$. Two
contacts and one barrier within the interacting part 
imply $G \sim T^{2 \alpha_\ix{B}}$ (see Fig.~\ref{fig:int_incommens_exp}).

In addition, in Fig.~\ref{fig:int_incommens_exp} we show
the effect of bulk contacts (in contrast to end contacts discussed so far).
Also for this setup $G(T)$ at sufficiently large $T$ is given approximately by 
$G \sim \sqrt{G_1G_2G_3}$. Since bulk contacts do not change significantly the
density of states in the wire \cite{comment_2}, their contribution
to the exponent in Eq.~\eqref{eq:phase_average_scaling} is $\alpha_i=0$.
This provides a concrete experimental setup to measure fingerprints of
phase averaging by considering a quantum wire with a single impurity
contacted by two end contacts and, in situ, also by two scanning
tunneling microscopy contacts situated left and right to the impurity
but far away from the end contacts compared to $L_T$. By either
measuring transport through the two end contacts, or one end and one
bulk contact, or two bulk contact, the situation (c), (d), and (e) of
Fig.~\ref{fig:int_incommens_exp} are realized, respectively. If phase
averaging dominates different exponents will be observed, whereas in
the case of dominant dephasing the exponent will always be given by
the impurity exponent $2\alpha_B$.\cite{temperature} 

An additional feature in Fig.~\ref{fig:int_incommens} is the occurrence
of peaks in $G(T)$ at certain moderate temperatures. They arise
when left- and right-dot resonances coincide leading to an avoided crossing
with a strongly enhanced total transmission.
As the transmission and reflection amplitudes of the single effective
barriers depend on $T$ via the range $L_T$ of the oscillations of the
effective potential, the dot resonances shift energetically with $T$.
The peak shifts occur in two steps, when
temperature is lowered from far larger than the bandwidth to zero: The first
step occurs when $T$ sweeps over the largest part of the band and is
due to a renormalization of the hopping in the chain. The second step
occurs when $T$ sweeps over the very peak under consideration. The
size in energy of this second step is proportional to the level
spacing of the corresponding dot and therefore in general different
for left- and right-dot resonances making possible an avoided crossing (see
Fig.~\ref{fig:int_incommens}).
At higher temperatures, there are so many transmission resonances
contributing to the conductance that the enlargement of single peaks
due to their avoided crossing is not visible in the conductance.

The temperature dependence of the dot resonances has also the
consequence that only two dots of exactly the same length
represent a truly commensurate configuration because
only for $L_\ix{L} = L_\ix{R}$ regularly coinciding left- and right-dot
resonances occur at every temperature. In this special case our data confirms that
Eq.~\eqref{eq:comm_cond}, which now reads $G^{-1} = G_1^{-1} + G_3^{-1}$,
is generalizable to interacting QWs. Depending on whether the
single barriers are within the wire or at the contacts,
this means $G \sim T^{2 \alpha_\ix{B}}$, $G\sim
T^{\alpha_\ix{B}}$ or a combination of both.

\section{Phase averaging for a wire with four barriers}
\label{sec:4barrier}

Finally we mention the case of four
barriers. Eq.~\eqref{eq:phase_average} then leads to 
\begin{equation}
  \label{eq:4barrier}
  \langle |t|^2 \rangle 
  =
  \frac{\sqrt{T_1 T_2 T_3 T_4}}{\agm{\sqrt{V+W}}{\sqrt{V-W}}},
\end{equation}
where $T_i = \abs{t_i}^2$, $W = 8 \abs{r_1 r_2 r_3 r_4}$,
\begin{equation}
  \label{eq:V}
  V 
  =
  8 \prod_i \left(1-\frac{1}{2}T_i\right) + \left(\frac{1}{2} + \sum_i
    \frac{1-T_i}{T_i^2} \right)\prod_i T_i
\end{equation}
and the arithmetic-geometric mean
$\agm{a}{b}$ of $a$ and $b$ is defined as the limit of the sequence
given by $a_0 = a$, $b_0 = b$, $a_{n+1} = \frac{1}{2}(a_n + b_n)$,
$b_{n+1} = \sqrt{a_n b_n}$.  For weak barriers, $T_i \approx 1$, this gives
again the product law $\langle |t|^2 \rangle \approx T_1 T_2 T_3 T_4$.
For strong barriers, $T_i\ll 1$, we get
\begin{equation}
  \label{eq:4barrier_strong}
  \langle \abs{t}^2 \rangle  \approx
  \frac{1}{2\pi} \sqrt{T_1 T_2 T_3 T_4}
  \ln \frac{16}{\sqrt{\sum_i T_i^2 + T_1 T_2 T_3 T_4 
      \sum_i \frac{1}{T_i^2}}}.
\end{equation}
The logarithmic factor 
appearing now causes small deviations from exact power-laws of $G(T)$. 
However, as for an increased number of barriers higher temperatures
are necessary for phase averaging to become a reasonable approximation,
this formula might not be as relevant as in the case of three barriers.

\section{Summary}
\label{sec:summary}
We have compared the effects of phase
averaging and dephasing in an one-dimensional mesoscopic system.
Investigating the scaling behavior of the conductance we have 
proposed that the two phenomena can be distinguished
experimentally by analyzing different kinds and numbers of barriers.
This is an important issue for the interpretation of past and future
experiments, especially due to the fact that also the contacts to
the leads form barriers. In cases where phase averaging
dominates, those contact barriers will influence the scaling behavior
even if their resistance is lower than the one of barriers within the
wire. This might be connected with still unexplained power-law
exponents in  transport measurements through carbon
nanotubes.\cite{postma,LL_theory,Tilman} 
Whether phase averaging or dephasing is dominant in these experiments
is still an open question and needs further investigations.

\section*{Acknowledgments}

The authors thank M.~B\"uttiker, S.~De~Franceschi, Y.~Gefen, Y.~Imry, 
F.~Reininghaus and K.~Sch\"on\-ham\-mer
for helpful discussions. This work has been supported by the
VW Foundation and the Forschungszentrum J\"ulich via the virtual 
institute IFMIT (S.J. and H.S.), the Deutsche 
Forschungsgemeinschaft via SFB 602 (V.M.), and 
the Alexander von Humboldt foundation and 
the INFM--SMC--CNR (T.E.).



\begin{thebibliography}{99}
\bibitem{comment_1} 
For strong hopping barriers, the reflection phases $\delta_i^\pm$
contribute $2ka$ to $\varphi_i$, so that the condition of
commensurability is to be formulated with $L_i + a$ or $N_i+1$ instead
of $L_i$ or $N_i$.
\bibitem{comment_2} 
We have neglected the contribution from the scaling
of the bulk density of states since its power-law exponent
is of second order in $U$ and therefore usually
smaller than the boundary one.
\end{thebibliography}
\end{document}